\documentclass[twocolumn,trackchanges]{aastex631}

\usepackage{graphicx, rotating,booktabs}
\usepackage{hyperref}

\begin{document}

\title{Vetting Ambiguous Planetary Candidates (APC) in TESS Data: Insights from LATTE Package and Detection of Nine Potential Exoplanet Candidates}

\correspondingauthor{Umer Hameed Khan}
\email{umer.spsc@gmail.com}

\author[0000-0002-1807-8864]{Umer Hameed Khan}
\affiliation{Pakistan Space and Upper Atmosphere Research Commission (SUPARCO),\\ SUPARCO Rd, P. O. Box No. 8402, Karachi 75270, Pakistan}

\author[0009-0006-8767-9379]{Khushlim Khan}
\affiliation{Pakistan Space and Upper Atmosphere Research Commission (SUPARCO),\\ SUPARCO Rd, P. O. Box No. 8402, Karachi 75270, Pakistan}

\author[0009-0000-5942-1764]{Muhammad Zeshan Ashraf}
\affiliation{Institute of Frontier and Interdisciplinary Science, Shandong University Qingdao, China}

\author[0009-0006-8767-9379]{A. Salam Khumber}
\affiliation{Pakistan Space and Upper Atmosphere Research Commission (SUPARCO),\\ SUPARCO Rd, P. O. Box No. 8402, Karachi 75270, Pakistan}

\begin{abstract}

We present the results of our study on ambiguous planetary candidates (APC) in the \textit{Transiting Exoplanet Survey Satellite (TESS)} data using the open source LATTE package. This tool can effectively distinguish transit-like signals, including planet candidates and eclipsing binaries, from stellar variability and instrumental noise. We conducted a comprehensive set of tests on the \textit{TESS} project candidates and visually inspected the resulting plots to filter out false alarms and false positives. Our study confirms the presence of 9 potential exoplanet candidates among the APCs, based on the lightcurve processing and subsequent analysis. These candidates are suitable for follow-up observations by ground-based observatories. Our results demonstrate the effectiveness of the LATTE package for the identification and vetting of exoplanet candidates in \textit{TESS} data. As new data from other \textit{TESS} sectors become available, we plan to extend our study to further refine our results and identify additional exoplanet candidates. 

\end{abstract}

\keywords{Astronomy data analysis(1858) --- Exoplanet astronomy(486) --- Exoplanet detection methods(489) --- Exoplanet catalogs(488)}

\section{Introduction} \label{sec:intro}
Detection of exoplanets has been a topic of intense interest and research in recent years. Over the past few decades, several methods have been developed and employed to detect exoplanets, ranging from the radial velocity method to the transit method and more \citep{mayor1995jupiter, bennett1996detecting, marois2008direct}.

Astronomy has recently paid a lot of attention to the application of machine learning (ML) techniques for the discovery and evaluation of exoplanets. The number of exoplanet candidates has rapidly expanded with the introduction of large-scale surveys like Kepler \citep{borucki2010kepler} and TESS \citep{ricker2015transiting}, necessitating the development of effective and precise algorithms for finding real planets among false positives. Several classical machine learning methods have been studied and implemented to classify light curves. These methods include decision trees, k-nearest neighbors, and random forests. These machine-learning methods have been applied to classify light curves in various studies, such as the Autovetter \citep{mccauliff2015automatic}, dimensionality reduction, and k-nearest neighbors \citep{thompson2015machine}, self-organizing maps (SOMs) \citep{armstrong2016transit}, the Robovetter project \citep{coughlin2016planetary}, and machine learning algorithm for the Next-Generation Transit Survey (NGTS) \citep{armstrong2018automatic}.

\cite{shallue2018identifying} conducted one of the early research utilizing machine learning (ML) for exoplanet discovery. They created a convolutional neural network (\textit{Astronet}) to recognize exoplanets in Kepler light curves. Their method identified exoplanets with a 98.8\%  accuracy and found a number of new exoplanets that were missed by earlier techniques. In their seminal paper, \citep{pearson2018searching} utilized deep neural networks that were trained on a large dataset of transits and non-transits, allowing the neural network to learn and distinguish between the two based on patterns in the data. Astronet was extended and improved by \citep{ansdell2018scientific}, who included additional scientific domain knowledge in the network architecture and input representations, resulting in a performance increase to 97.5\% accuracy and 98.0\% average precision. \citep{dattilo2019identifying} utilized the Astronet model for exoplanet detection in the Kepler K2 dataset (\textit{Astronet-K2}), which resulted in the identification of two new super-Earth exoplanets. \cite{yu2019identifying} and \citep{osborn2020rapid} reported the adaptation and modification of the Astronet neural network to analyze data from the \textit{TESS} mission. The Astronet was specifically tailored to address the unique characteristics of the \textit{TESS} data, enabling efficient and accurate automated triage and vetting of exoplanet candidates.
More recent contributions are made by \citep{jara2020transiting, armstrong2021exoplanet, rao2021nigraha, malik2022exoplanet, fernandes2022pterodactyls, valizadegan2022exominer, magliano2023tess}. 

Here, we have used the LATTE: Lightcurve Analysis Tool for Transiting Exoplanets \citep{eisner2020latte} package for the identification and vetting of exoplanet candidates in \textit{TESS} data. The LATTE Python package is a valuable resource for analyzing TESS lightcurves, as it provides a suite of standard diagnostic tests that aid in identifying, verifying, and characterizing signals. Section \ref{sec:data} outlines the data and procedures for vetting the candidates, and Section \ref{sec:tests} details the diagnostic tests performed on the candidates to determine their authenticity. Finally, in Section \ref{sec:discu}, we summarize our findings and their significance for exoplanet research.

\section{Data Set} \label{sec:data}
TESS two-minute cadence data from recent sectors was used to identify exoplanet candidates. The data was sourced from the NASA Exoplanet Archive \footnote{\url{https://exoplanetarchive.ipac.caltech.edu/}}, which is a reliable database of confirmed exoplanets, planetary candidates, and their parameters.  The TESS Follow-Up Observing Program (TFOP) categorizes potential exoplanet candidates detected by TESS based on their characteristics. Dispositions include False Positive (FP): Non-planetary sources with transit-like signals, Planet Candidate (PC): TCEs likely to be genuine exoplanet candidates, Confirmed Planet (CP): Exoplanet candidates confirmed through independent observations, Known Planet (KP): Candidates corresponding to known exoplanets or transiting systems, and Ambiguous
Planetary Candidate (APC): candidates requiring further investigation due to uncertain features. APCs may arise from instrumental effects, stellar activity, or other phenomena. Additional observations are needed to distinguish genuine planetary transits from false positives. 

The classification of Ambiguous Planetary Candidates (APCs) presents a notable challenge and necessitates further investigation to establish their true nature. In this study, we aim to address this ambiguity by selecting APCs as our focal point and subjecting them to comprehensive vetting checks using the LATTE package. Our primary objective is to capitalize on the capabilities of LATTE to conduct rigorous analyses and diagnostics, including background flux examination, in-out transit flux assessment, pixel-level light curve analysis, and centroid correlation/positions evaluation for each individual transit event. These checks enable us to attain a clearer understanding of the APCs, either confirming their authenticity as genuine planet candidates or identifying them as false positives.
We filtered out ambiguous planetary candidates (APCs) from the TESS Project Candidates (TPC). Out of the 6,213 TPCs, only 373 candidates were categorized as APCs. We obtained TESS project data for the APCs that had already passed initial vetting criteria, including their brightness, transit depth, and transit duration. We chose these APCs because they had not yet been confirmed or rejected by previous studies, making them suitable for further investigation. We downloaded the lightcurves of these candidates in the TESS short cadence data from the Mikulski Archive For Space Telescopes (MAST) portal \footnote{\url{https://mast.stsci.edu/portal/Mashup/Clients/Mast/Portal.html}}, which provides high precision and rapid observations of a given target. Only 268 of these candidates had 2-minute cadence data available, comprising the final list of exoplanet candidates analyzed in this study.

\subsection{Methodology}
LATTE \citep{eisner2020latte} offers a comprehensive set of diagnostic tests that can be applied to TESS lightcurves for the identification and vetting of signals accurately. These tests include momentum dumps, target properties tables, background flux plots, and x and y centroid positions, which are essential for identifying and removing false positives caused by various factors, including instrument systematics and astrophysical variability.

The program is intended to conduct fast and thorough analyses of promising candidates previously identified through different methods such as citizen science, and TESS pipelines as mentioned in \citep{fischer2012planet, christiansen2018k2, eisner2020planet}. The tool conducts diagnostic tests similar to data validation procedures performed by Science Processing Operations Center and Quick-Look Pipeline for short and long-cadence data respectively, as described by \citep{huang2019quick}.

However, LATTE extends the scope of analysis to any TESS target with a valid TIC ID, including both two-minute and thirty-minute cadence targets \citep{stassun2019revised}. While the SPOC pipeline tests are limited to a selection of two-minute cadence TESS targets only, LATTE provides a more comprehensive analysis of all TESS targets. The program's ability to analyze a broader range of targets with detailed diagnostic tests enables it to obtain more detailed insights and increases the potential for identifying promising candidates for further study. It has a fast analysis capability that allows for the rapid evaluation of a larger number of targets, ultimately facilitating the search for new discoveries in the field of astronomy.

\section{Light-curve-based vetting checks} \label{sec:tests}
The Lightcurve Analysis Tool for Transiting Exoplanet LATTE package was used to perform several diagnostic tests on the 268 APCs. The results of these tests for TIC 270515566 are summarized below:

\begin{figure}[ht!]
    \centering
    \includegraphics[width=\columnwidth]{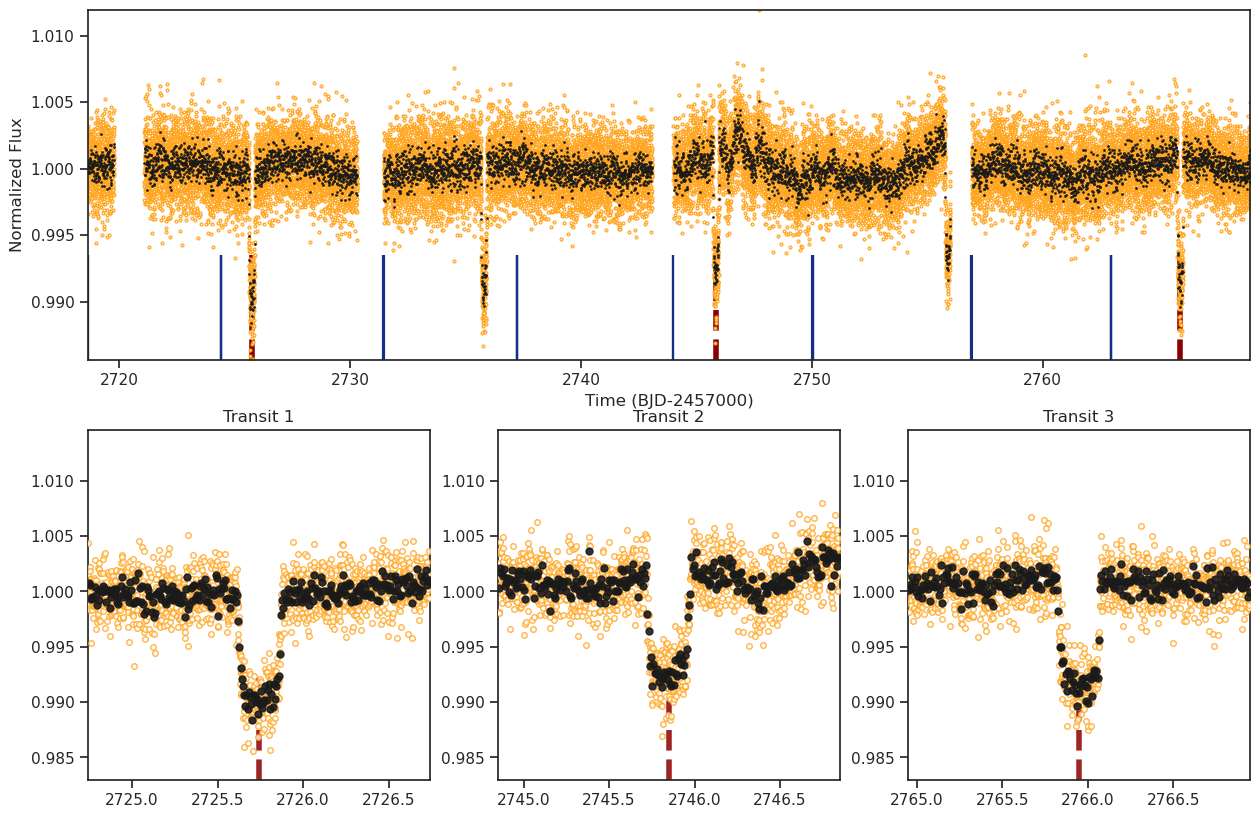}
    \caption{Full lightcurve for TIC 270515566 with marked transit events (red dashed lines) and reaction wheel momentum dumps (solid blue lines). Momentum dumps occurred every 2-2.5 days and lasted about a couple of hours.}
    \label{fig:01}
\end{figure}

\subsection{Background Flux}
TESS's background flux varies due to its eccentric orbit and lunar resonance, but the TESS pipeline removes most of these effects. However, sudden changes in background flux caused by passing objects can introduce spurious signals into the light curve. Careful inspection of background plots is necessary to identify such contamination. Other sources of contamination may also affect the light curve, requiring additional processing and analysis techniques for reliable transit detection.

\begin{figure}[ht!]
    \centering
    \includegraphics[width=\columnwidth]{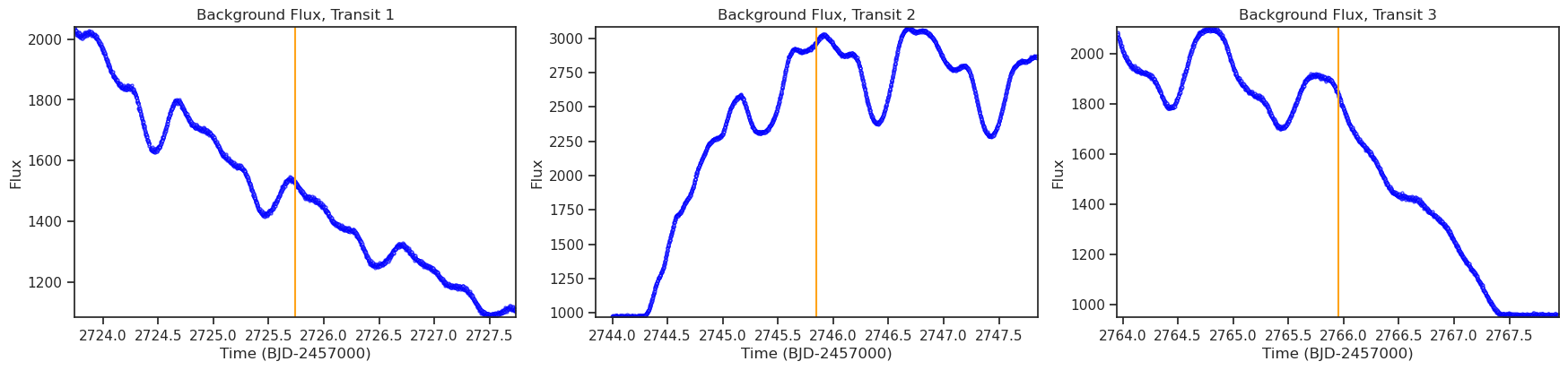}
    \caption{Background flux as a function of time, where the time of the transit-like event is marked by an orange vertical line.}
    \label{fig:02}
\end{figure}

\subsection{Centroid Positions}
In order to maintain the observational stability of the Transiting Exoplanet Survey Satellite (TESS), we employ a monitoring strategy focused on the positional stability of the weighted average of pixel values within the aperture over time. This meticulous analysis allows us to identify any sudden spikes or significant variability coinciding with transit-like events, which can serve as indicators of systematic errors affecting the observed light curve.

By adopting this approach, we gain an additional diagnostic tool to detect and mitigate systematic effects that may potentially impact the accuracy of transit detections. The continuous monitoring of the aperture’s positional stability enables us to gain a comprehensive
understanding of instrumental and environmental influences, thus facilitating the implementation of appropriate corrective measures. The assessment of the x and y positions of the aperture’s weighted average pixel values provides crucial insights into any spurious signals that may arise due to various sources of noise or biases. Through careful examination and interpretation of the observed positional changes, we can discern between genuine transit signals and undesired artifacts.

\begin{figure}[ht!]
    \centering
    \includegraphics[width=\columnwidth]{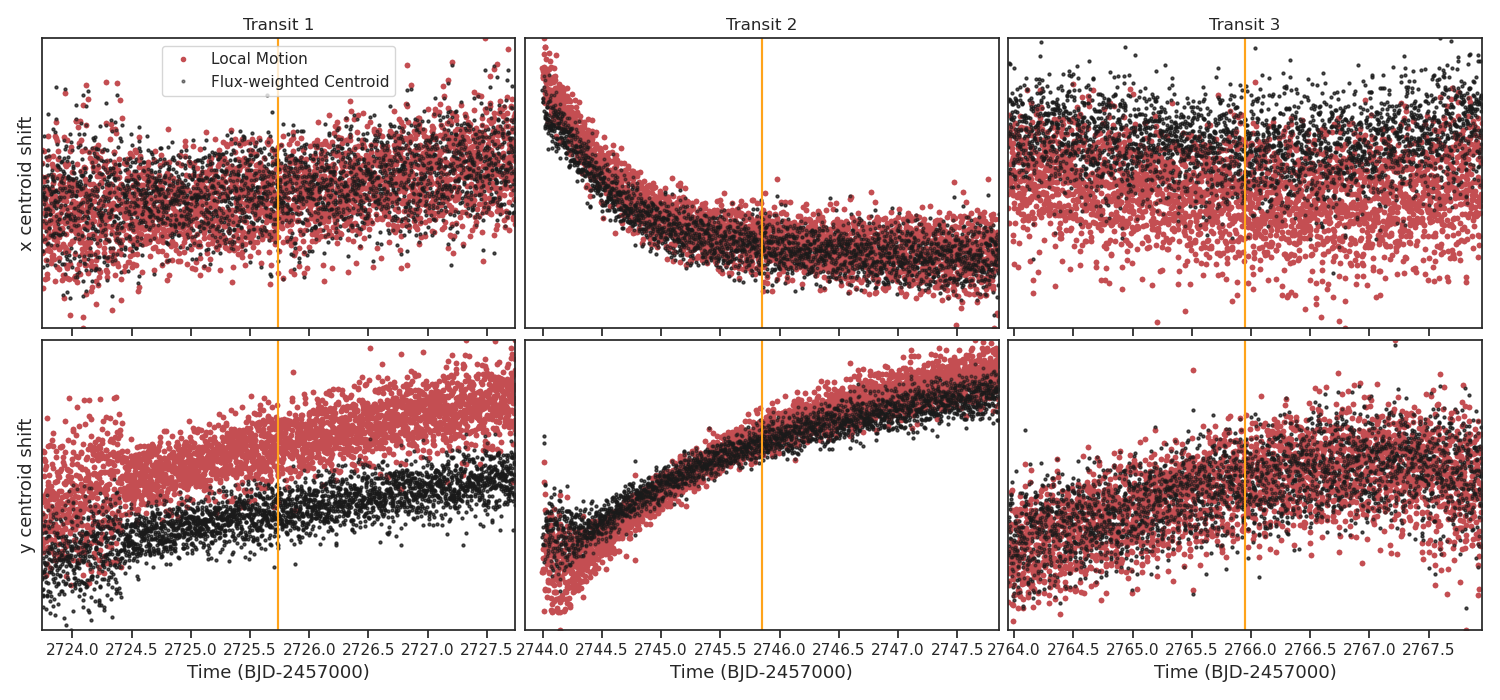}
    \caption{The x and y centroid positions of the target's flux-weighted centroid around the time of each transit-like event are shown by the black points on the CCD. The red points indicate the local motion differential velocity aberration (DVA), pointing drift, and thermal effects on the CCD column and row. The vertical orange line denotes the time of the transit-like event.}
    \label{fig:03}
\end{figure}

\subsection{Aperture Size}
The constancy of transit shape and depth across different extraction apertures is a key characteristic of planet-induced transits occurring in stellar systems. However, variations in shape or depth observed when using different aperture sizes may be attributed to the presence of a halo-effect caused by reflections within the stellar "halo." Therefore, a precise selection of extraction apertures plays a critical role in ensuring the reliable detection of exoplanetary systems, while diligent consideration of potential contamination sources is essential for achieving high precision.

The accurate determination of extraction apertures is crucial as it directly impacts the fidelity of transit signal detection. A systematic and careful approach to selecting the optimal aperture size minimizes the introduction of unintended biases and reduces the risk of false-positive or false-negative results. Additionally, accounting for potential contamination sources, such as background
stars or instrumental artifacts are of utmost importance to enhance the precision of transit detection. Lightcurves in Figure 4 were obtained by extracting photometric measurements using different aperture sizes. The consistency of the transit shape and depth across varying aperture sizes provides compelling evidence of a genuine planetary transit event occurring within the observed stellar system. This observation suggests that the underlying transit signal remains robust and unaffected by changes in the aperture size employed during the data extraction process.

\begin{figure}[ht!]
    \centering
    \includegraphics[width=\columnwidth]{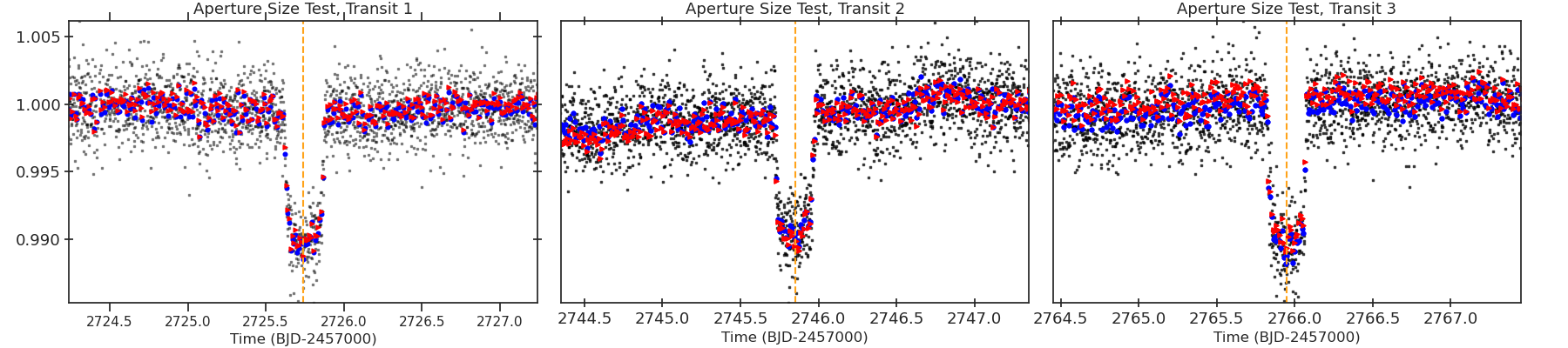}
    \caption{The LC extracted using different aperture sizes. The transit shape and depth are the same with both apertures making it a good planet candidate.}
    \label{fig:04}
\end{figure}

\subsection{Aperture Masks}
Transit-like events resulting from a blended eclipsing binary system may exhibit differences in transit depth and shape between two distinct light curves. Accurately identifying such signals is crucial for characterizing exoplanetary systems and their host stars. However, a sub-optimal selection of masks used to extract the transit signal can lead to the misinterpretation of a moving signal. To mitigate this, it is important to ensure that the masks used in the analysis accurately target the desired region and have appropriate sizes and shapes. A careful evaluation of the light curves and masks is therefore essential for identifying and characterizing transit-like events resulting from blended eclipsing binaries.

\begin{figure}[ht!]
    \centering
    \includegraphics[width=\columnwidth]{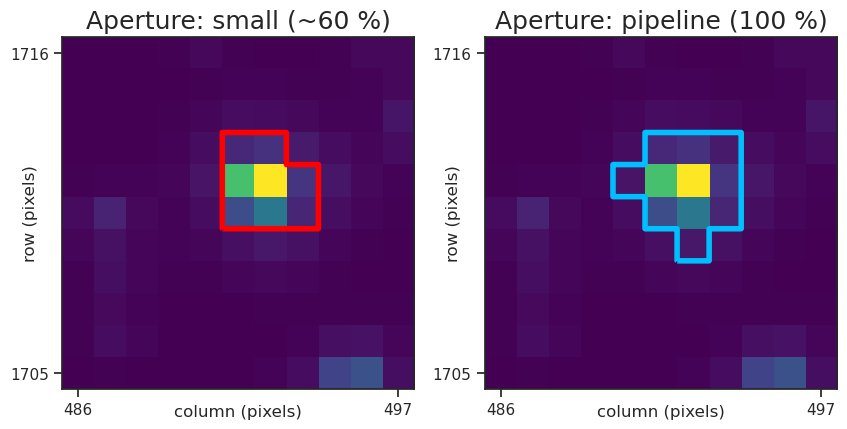}
    \caption{Aperture masks with two different sizes for TIC 270515566. The apertures in both cases are accurately centered on the target, denoted by the brightest pixel, enabling effective use of the aperture size test to determine whether the observed signal results from a transiting exoplanet.}
    \label{fig:05}
\end{figure}

\subsection{In and out of transit average flux}
The comparison of the average flux during the transit-like event with that outside of it is performed to identify any significant differences. The left and middle images represent the average flux during and outside the transit event, respectively. The average in-transit and average out-of-transit flux values are calculated by considering the flux measurements extracted from the Target Pixel Files (TPFs). These flux values are corrected for systematic effects using techniques such as Principal Component Analysis (PCA) as outlined in \citep{stumpe2012kepler}.

During the vetting process, the average in-transit flux is obtained by calculating the mean flux value during the transit event, typically defined by the transit duration. The transit duration can vary depending on the specific parameters of the exoplanetary system under investigation. On the other hand, the average out-of-transit flux is calculated by determining the mean flux value outside of the transit event, typically in a window of time before or after the transit. The primary focus is on the right image which is the difference between the two. If there are any changes in the average flux in the right image, such as a shift in the location of the brightest pixel, it indicates that the change in brightness is likely caused by an object other than the star under observation. For instance, it could be due to a background eclipsing binary.

\begin{figure}[ht!]
    \centering
    \includegraphics[width=\columnwidth]{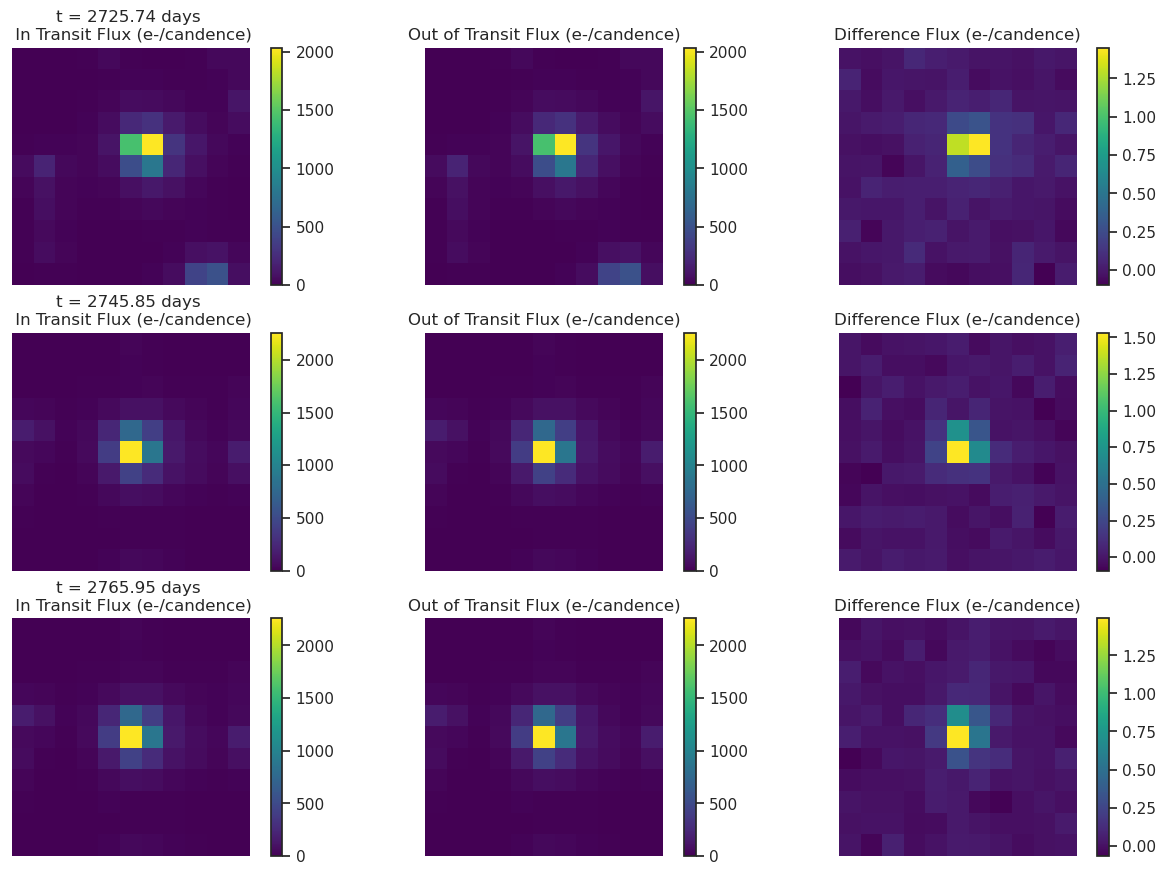}
    \caption{This figure shows the different images for target TIC 270515566 during each transit-like event. The left image displays the average in-transit flux, the middle image displays the average out-of-transit flux (mean flux value during the transit event typically defined by the transit duration), and the right image displays the difference between the two.}
    \label{fig:06}
\end{figure}

\subsection{LCs of nearby stars}
This test involves examining the lightcurves of five stars closest in proximity to our target star, all of which were observed by TESS every two minutes. If a similar dip is observed in the lightcurves of these nearby stars at the same time as the transit-like event in our candidate, it is likely that the signal is caused by a different star or a background event. However, the identification of the actual star responsible for a transit event depends on various factors such as distances, magnitude differences, contamination ratios, and the presence of other stars in the vicinity of the observed star.

In some cases, if a transit signal appears stronger on a nearby star compared to the target star, it may suggest that the nearby star is the likely source. However, even in such instances, it’s important to consider other factors before drawing a definitive conclusion. Factors like the relative distances between the stars, the difference in their magnitudes, and the possibility of contamination from neighboring stars must be carefully assessed. Furthermore, the presence of other stars near the observed star can significantly complicate the situation. These neighboring stars may contribute to the observed transit signal, leading to confusion in determining the true source. Additionally, variations in the brightness of the observed star caused by factors such as stellar activity or instrumental artifacts need to be considered and carefully distinguished from transit signals. 

\begin{figure}[ht!]
    \centering
    \includegraphics[width=\columnwidth]{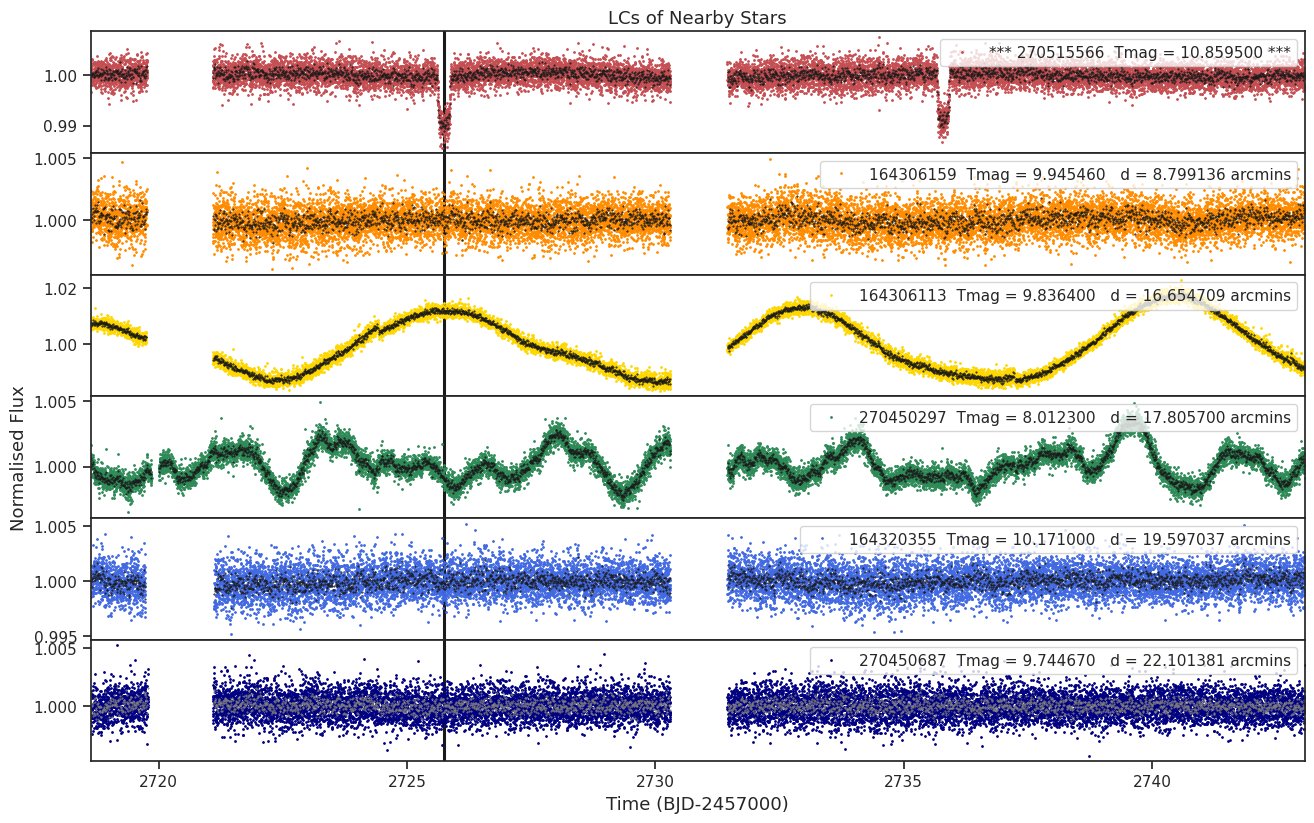}
    \caption{The lightcurves of the five stars closest to our target, 270515566, are displayed in the top panel. The distances between each star and the target, as well as their respective TESS magnitudes, are also provided.}
    \label{fig:07}
\end{figure}

\subsection{Pixel level lightcurve plot}
In order to ascertain the precise localization of the transit event and differentiate it from other regions within the image, we have conducted a pixel-level analysis. This approach provides a comprehensive examination of each individual pixel in the vicinity of the
target star, enabling us to accurately attribute the transit event to the intended target.

To illustrate our findings, we present a plot depicting the light curves of every pixel surrounding the target star. Within this plot, the transit event is denoted by red and yellow data points, drawing attention to the temporal changes in brightness associated with the
transit phenomenon. However, it is important to note that the individual plots within the overall visualization are small, posing a challenge in identifying shallow transit events that exhibit subtle variations in flux. Nonetheless, through this pixel-level analysis, we ensure a thorough investigation into the origin of the transit event. By scrutinizing the light curves of individual pixels surrounding the target star, we are able to assess the unique contributions of each pixel and discern any deviations that may be indicative of transit activity. This rigorous examination plays a crucial role in confirming the localization of the transit event and ruling out any alternative sources within the image.

\begin{figure}[ht!]
    \centering
    \includegraphics[width=\columnwidth]{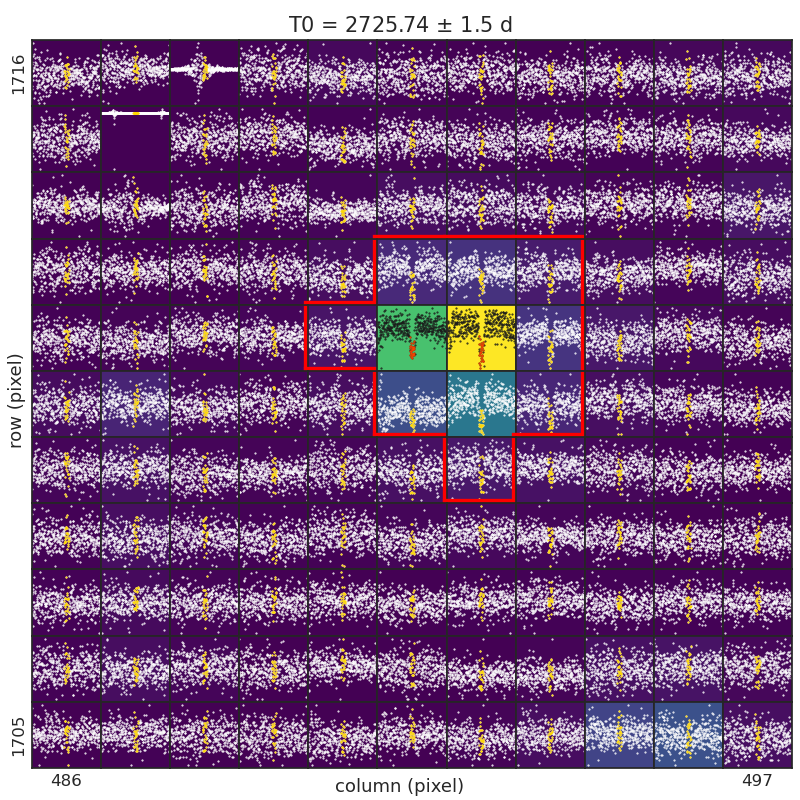}
    \caption{The normalised flux for each pixel around the time of the transit-like event was extracted using the SPOC pipeline mask. The in-transit data is indicated by the orange data points, while the solid red lines represent the SPOC pipeline mask. The figure shows the results for one sector only.}
    \label{fig:08}
\end{figure}

\begin{table*}[ht!]
    \centering
    \caption\\
    {Table containing the TIC ID, most recent available short cadence TESS sectors, Right ascension, Declination, TESS magnitude, Planet orbital period, planet radius, planet transit depth, and transit duration.}
    
    \begin{tabular}{|c|c|c|c|c|c|c|c|c|c|}
    \toprule
    \hline
         TIC & Sector & RA & Dec & Tmag & Orb Period & Radius & Transit Depth & Transit Duration\\
         & & (deg)& (deg)& (mag) & (days) & (R\_Earth) & (ppm) & (hours)\\ \hline
         270515566 & 52,53 & 264.454 & 47.23 & 10.86 & 10.05 & 13.59 & 8660 & 6.13\\
         350743714 & 12,13 & 88.886 & -57.29 & 9.67 & 7.76 & 12.92 & 7636 & 3.25\\
         290036495 & 46 &192.66 & -14.57 & 12.11 & 5.25 & 20.78 & 23400 & 1.83\\
         180652891 & 49 & 199.61 & 35.95 & 9.24 & 1.55 & 10.42 & 2450 & 1.14\\
         123898871 & 33 & 90.92 & -19.04 & 9.85 & 4.91 & 20.09 & 12846 & 4.13\\
         115771549 & 56 & 339.82 & 33.89 & 8.38 & 4.72 & 15.28 & 3000 & 4.05\\
         100097716 & 29,30 & 24.05 & -42.30 & 11.35 & 3.07 & 25.20 & 8545 & 1.70\\
         71612483 & 49,50 & 202.29 & 31.40 & 11.495 & 1.22 & 16.02 & 15450 & 1.85\\ 
         67196573 & 33 & 89.46 & -11.42 & 10.79 & 2.56 & 17.79 & 1860 & 1.05\\
         61980227 & 28 & 335.54 & -30.76 & 12.05 & 3.10 & 19.24 & 10354 & 1.86\\
    \hline     
    \end{tabular}
    \label{table:01}
\end{table*} 

\subsection{Checking for the contaminant transiting source}
For accurate identification of the transit source, we have used the open source python package Contaminante \footnote{\url{https://github.com/christinahedges/contaminante}}. The Contaminante package is a valuable tool for researchers working with NASA's Kepler, K2, or TESS data to identify potential contaminant sources in their observations. When searching for transiting planets, it is essential to distinguish between signals from actual exoplanets and those that may be caused by neighboring contaminants, such as eclipsing binaries or background stars.

Contaminante provides a user-friendly interface for analyzing light curves and identifying potential contaminant sources in the data. By comparing the transit signal to the surrounding light curve, users can determine whether the signal is consistent with a transiting exoplanet or whether it is likely caused by a neighboring contaminant. We incorporated the Contaminant package as an additional verification step prior to the final vetting of our candidates. In our investigation, we subjected 18 candidate signals to rigorous scrutiny using the Contaminant package. This thorough analysis allowed us to confirm that the transiting source in 10 candidates indeed originated from the intended target, ruling out the presence of extraneous contaminating sources within the data. This confirmation provided strong evidence for the existence of genuine exoplanets in these cases. However, our analysis also revealed contaminant sources in the remaining 8 candidates. These findings led us to exclude these candidates as false positives, as the signals were determined to be inconsistent with the expected characteristics of transiting exoplanets. This demonstrates the effectiveness of the Contaminant package in identifying and mitigating potential sources of contamination, thereby ensuring the reliability and accuracy of our results. This finding underscores the importance of implementing robust techniques for differentiating between true planetary signals and spurious contaminants in the pursuit of exoplanet identification.

In order to establish the threshold values used in the vetting tests for identifying potential exoplanet candidates, we employed a combination of established algorithms and meticulous analysis. Our approach involved applying the vetting checks of LATTE to confirmed and known planets within the data obtained from the Transiting Exoplanet Survey Satellite (TESS). This initial step allowed us to identify common patterns and characteristics that are indicative of genuine exoplanet signals.

By closely examining the behavior of these confirmed planets, we were able to establish a baseline for comparison. This baseline served as a reference to discern the features that reliably indicate the presence of an exoplanet. We carefully analyzed various parameters such as the transit depth, transit duration, and transit shape to determine the typical values exhibited by confirmed exoplanets. These parameters are critical in assessing the likelihood of a candidate being a genuine exoplanet.

Furthermore, we conducted a visual comparison of individual plots of the APC signals with those of confirmed planets. This involved a side-by-side examination of the APC plots and the plots of known exoplanets. This visual analysis played a crucial role in differentiating between potential exoplanet candidates and false alarms or false positives. By carefully scrutinizing the shape, periodicity, and other visual characteristics of the signals, we could identify potential candidates that resembled the behavior of confirmed exoplanets.
After conducting rigorous vetting tests, we have successfully recovered ten potential exoplanet candidates from the APCs. These candidates have undergone a thorough screening process to assess their viability and authenticity as exoplanets. To provide a comprehensive overview of our findings, a detailed summary of each candidate is presented in Table ~\ref{table:01}.

\begin{figure}[ht!]
    \centering
    \includegraphics[width=\columnwidth]{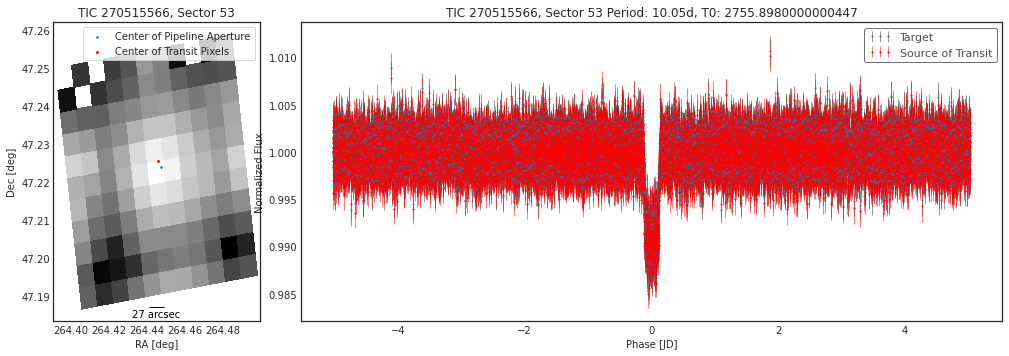}
    \caption{The plot displays a transit that is aligned with the target, providing evidence that the transiting signal originates from the target itself and not from any nearby contaminants.}
    \label{fig:09}
\end{figure}

\section{Discussion and Conclusion} \label{sec:discu}
The search for exoplanets beyond our solar system has witnessed remarkable advancements, driven by space missions such as the Transiting Exoplanet Survey Satellite (TESS). This ongoing pursuit has yielded a vast number of potential candidates, each necessitating meticulous analysis and verification to establish their true nature. In this context, the Ambiguous Planetary Candidates (APCs) emerge as an intriguing avenue for vetting and confirming exoplanet candidates that have been flagged as ambiguous by the TESS Follow-Up Observing Program (TFOP).

To comprehensively assess the viability of these candidates, our study employed the sophisticated LATTE (Lightcurve Analysis for Transiting Exoplanets) package on a sizeable sample of 268 objects sourced from the TESS project candidates list. The LATTE program is a comprehensive analysis tool developed to thoroughly investigate the transit events observed by the Transiting Exoplanet Survey Satellite (TESS). By examining all the transit events detected by TESS, LATTE aims to eliminate false alarms and identify any potential instrumental errors that may affect the accuracy of transit measurements.

Drawing inspiration from the SPOC pipeline data validation, LATTE employs multiple rigorous tests to evaluate various aspects of the transit events. These tests include assessing the background flux, in-out transit flux, pixel-level light curves, and centroid correlation/positions for each individual transit. By scrutinizing each transit event separately, LATTE ensures a meticulous analysis and allows for the identification and mitigation of any false alarms or instrumental artifacts that might impact the transit data.

To rule out the possibility of a transit-like event being caused by a background object, such as a solar system object or asteroid in the line of sight, LATTE employs comparative plots of background flux and overall flux. During TESS’s perigee, when it reaches the closest point to Earth in its eccentric orbit, enhanced scattered light in the telescope optics can cause a significant increase in background flux.

Hence, LATTE carefully examines the background plots to verify that no sudden spikes occur precisely at the time of the transit-like events. This comprehensive analysis and scrutiny of the transit events through the LATTE program demonstrate its efficacy in ensuring the accuracy and reliability of the TESS observations. The utilization of comparative plots and the consideration of background flux fluctuations provide valuable insights for distinguishing genuine transit signals from potential false alarms caused by instrumental effects or background objects. This approach allowed for the identification of 258 APCs that exhibited failed vetting tests, thus warranting their exclusion as false positives. However, among the candidates subjected to this stringent vetting process, ten candidates successfully
passed all tests, manifesting as promising exoplanet candidates worthy of further investigation and scrutiny,
particularly through ground-based observations.

Looking ahead, future observations employing advanced telescopes and cutting-edge instruments, such as high-resolution spectroscopy, hold immense potential for providing crucial insights into the intrinsic properties and potential habitability of these exoplanets. These endeavors aim to unravel the intricate details of these celestial bodies, shedding light on their atmospheric compositions, surface features, and potential for hosting life. Moreover, as the field of exoplanet research evolves, the development and implementation of innovative vetting methods, exemplified by the LATTE package utilized in our study, play a pivotal role in enhancing our understanding of the exoplanet population and refining our techniques for identifying genuine exoplanet candidates amidst the vast
sea of possibilities.

In conclusion, our study represents a significant stride forward in the pursuit of characterizing and validating
exoplanet candidates within the TESS data. Through a rigorous vetting process targeting the ambiguous planetary candidates (APCs), we have successfully recovered ten potential exoplanet candidates that demonstrate promise and merit further investigation. The precision and accuracy of our analysis were bolstered by meticulous evaluation of light curves, judicious selection of aperture masks, and vigilant monitoring of aperture positional stability to identify
and mitigate systematic effects. Additionally, the incorporation of advanced analysis techniques served to enhance the reliability of our exoplanet detection and minimize potential contamination sources. These findings underscore the
significance of robust vetting procedures in discerning and characterizing exoplanet candidates, particularly in
cases involving ambiguous candidates. By employing such comprehensive methodologies and continuing to refine our techniques, we advance our understanding of the exoplanet population, contributing to the broader exploration of our place in the cosmos.

\begin{acknowledgments}
\textit{We are thankful to the anonymous referee for their insightful comments, which have significantly enhanced the quality and clarity of the manuscript. The authors would like to extend their gratitude to the Space and Upper Atmosphere Research Commission (SUPARCO) for providing the opportunity to conduct this research. This research has made use of the NASA Exoplanet Archive, which is operated by the California Institute of Technology, under contract with the National Aeronautics and Space Administration under the Exoplanet Exploration Program. This paper includes data collected with the {\it TESS} mission, obtained from the {\it MAST} data archive at the Space Telescope Science Institute (STScI). Funding for the {\it TESS} mission is provided by the {\it NASA} Explorer Program. STScI is operated by the Association of Universities for Research in Astronomy, Inc., under {\it NASA} contract NAS 5–26555.}
\end{acknowledgments}

\software{LATTE \citep{eisner2020latte}, Lightkurve \citep{cardoso2018lightkurve}
        }

\bibliography{References}{}
\bibliographystyle{aasjournal}

\end{document}